**Scaling up Search Engine Audits: Practical Insights for Algorithm Auditing**

Roberto Ulloa[1], Mykola Makhortykh[2] and Aleksandra Urman[3]

[1] GESIS – Leibniz-Institute for the Social Sciences

[2] University of Bern

[3] University of Zurich

**Author Note**

Preprint and working version. We have no conflicts of interest to disclose. Data collections were sponsored by the SNF (100001CL_182630/1) and DFG (MA 2244/9-1) grants for the project "Reciprocal relations between populist radical-right attitudes and political information behaviour: A longitudinal study of attitude development in high-choice information environments" lead by Silke Adam (U of Bern) and Michaela Maier (U of Koblenz-Landau), and the FKMB (the Friends of the Institute of Communication and Media Science at the University of Bern) grant "Algorithmic curation of (political) information and its biases" awarded to Mykola Makhortykh and Aleksandra Urman.

**Abstract**

Algorithm audits have increased in recent years due to a growing need to independently assess the performance of automatically curated services that process, filter, and rank the large and dynamic amount of information available on the internet. Among several methodologies to perform such audits, virtual agents stand out because they offer the ability to perform systematic experiments, simulating human behaviour without the associated costs of recruiting participants. Motivated by the importance of research transparency and replicability of results, this paper focuses on the challenges of such an approach. It provides methodological details, recommendations, lessons learned, and limitations based on our experience of setting up experiments for eight search engines (including main, news, image and video sections) with hundreds of virtual agents placed in different regions. We demonstrate the successful performance of our research infrastructure across multiple data collections, with diverse experimental designs, and point to different changes and strategies that improve the quality of the method. We conclude that virtual agents are a promising venue for monitoring the performance of algorithms across long periods of time, and we hope that this paper can serve as a basis for further research in this area.

**Introduction**

The high and constantly growing volume of internet content creates a demand for automated mechanisms that help to process and curate information. However, by doing so, the resulting information filtering and ranking algorithms can steer individuals' beliefs and decisions in undesired directions [1–3]. At the same time, the dependency that society has developed on these algorithms, together with the lack of transparency of companies that control such algorithms, has increased the need for algorithmic auditing, the "process of investigating the functionality and impact of decision-making algorithms" [4]. A recent literature review on the subject identified 62 articles since 2012, 50 of those published between 2017 and 2020, indicating a growing interest from the research community in this method [5].

One of the most studied platforms are web search engines - almost half of the auditing works reviewed by Bandy (2021) were focused on Google alone - as a plethora of concerns have been raised about representation, biases, copyrights, transparency, and discrepancies in their outputs. Research has analyzed issues in areas such as elections [6–11], filter bubbles [12–17], personalized results [18, 19], gender and race biases [20–22], health [23–25], source concentration [10, 26–29], misinformation [30], historical information [31, 32] and dependency on user-generated content [33, 34].

Several methodologies are used to gather data for algorithmic auditing. The data collection approaches range from expert interviews to Application Programming Interfaces (APIs) to data donations to virtual agents. The latter refer to programs (scripts or routines) that simulate user behaviour to generate data outputs from other systems. In a review of auditing methodologies [35], the use of virtual agents (referred to as agent-based testing) stands out as a promising approach to overcome several limitations in terms of applicability, reliability, and external validity of audits, as it allows to systematically design experiments by simulating human behaviour in a controlled environment. Around 20% (12 of 62) of the works listed by Bandy (2021) use this method. Although some of the studies share their programming code to enable the reproducibility of their approach [18, 24, 36], the methodological details are only briefly summarized. The challenges involved in implementing the agents and the architecture that surround them are often overlooked. Thus, the motivation for this paper is to be transparent to the research community, allow for the replicability of our results, and transfer knowledge about lessons learned for the process of auditing algorithms with virtual agents.

Out-of-the-box solutions to collect data for algorithmic auditing do not exist because websites (and their HTML) evolve rapidly, so the data collection tools require constant updates and maintenance to adjust to these changes. The closest option to such an out-of-the-box solution is provided by Haim (2020), but even there the researcher is responsible for creating the necessary "recipes" for the platforms they audit - and these recipes will most certainly break as platforms evolve. Therefore, this work focuses on considerations, challenges, and potential pitfalls of the implementation of an infrastructure that systematically collects large volumes of data through virtual agents that interact with a specific class of platforms - that is search engines -, to give other researchers a broader perspective on the method.

To evaluate the performance of our method we pose two research question: RQ1: How are (a) data coverage and (b) effective size affected when audits are applied at a large scale? RQ2: What practical challenges emerge when scaling up search engine audits?

Our results demonstrate the success of our approach by often achieving near-perfect coverage and consistently collecting above 80% of results. Additionally, the overall effective size of the collection is above 95%, and we retrieved the exact number of pages in more than 92% of the cases. We use those "exact" cases to provide size estimates of search results, and we demonstrate that they can be used to successfully calculate data collection sizes using an out-of-sample approach. By providing disaggregate figures per collection and search sections, we also show how strategic interventions improve coverage in later rounds.

We provide a detailed methodological description of our approach including the simulated behaviour, the configuration of environments and the experimental designs for each collection. Additionally, we discuss contingencies included in our approach to cope with (a) personalization, the adjustment of search results according to the user characteristics such as location or browsing history, and (b) randomization, difference of search results that emerge even on the same browsing conditions of the audited systems' outputs. Both issues can distort the process of auditing if not addressed properly. For example, we synchronize the search routines of the agents and utilize multiple machines and different IP addresses under the same conditions to capture the variance of unknown sources. We focus on simulating user behaviour in controlled environments in which we manipulate several experimental factors: search query, language, type of search engine, browser preference and geographic location. We collect data coming from the text (also known as main or default results), news, image, and video search results of eight search engines representing the United States, Russia and China.

Our main contributions are the presentation of a comprehensive methodology for systematically collecting search engine results at a large scale, as well as recommendations and lessons learned during this process, which could lead to the implementation of an infrastructure for long-term monitoring of search engines. We demonstrate the successful performance of our research infrastructure across multiple data collections, and we provide average search section sizes that are useful to calculate the scale of future data collections.

The rest of this paper is organized as follows: section 2 discusses related work in the field of algorithmic auditing, in particular studies that have used virtual agents for data collection. Section 3 presents our methodology in terms of architecture and features of our agents, including detailed pitfalls and improvements in each stage of development. Section 4 presents the experiments corresponding to our data collection, and the response variables that are used to evaluate the performance of the method. Section 5 presents the results for each round of data collection according to browser, search engine and type of result (text, news, images and video). Section 6 discusses the achievements of our methodology, lessons learned from the last two years of research, and important considerations to successfully perform agent-based audits. Section 7 concludes with an invitation of scaling search engine audits further with long-term monitoring infrastructures.

**Related work**

Until now, the most common methodology to perform algorithm auditing is through APIs [5]. This approach is relatively simple because the researcher accesses clean data directly produced by the provider, avoiding the need to decode website layouts (represented in HTML). However, it ignores user behaviour (e.g. clicking, scrolls, loading times) as well as the environment in which behaviour takes place (e.g. browser type, location). For example, in the case of search engines, it has been shown that APIs sometimes return different results than standard webpages [37], and that results are highly volatile in general [38]. An alternative to using APIs is to recruit participants and collect their browsing data by asking them to install software (i.e. browser extensions) to directly observe their behaviour (e.g.,

Bodo et al., 2018; Möller et al., 2020; Puschmann, 2019; Robertson, Jiang, et al., 2018). Although this captures user behaviour in more realistic ways, it requires a diverse group of individuals who are willing to be tracked on the web and/or capable of installing tracking software on their machines [35, 41]. Additionally, it is difficult to systematically control for sources of variation, such as the exact time in which the data is accessed in the browser, and personalization, such as the agent's location. Compared with these two alternatives, virtual agents allow for flexibility to perform systematic experiments that include individual behaviour in realistic conditions, without the costs involved in recruitment of human participants.

Several studies have used virtual agents to conduct audits of search engine performance with a variety of criteria. Feutz et al. [42] analyzed changes in search personalization based on accumulation of data about user browsing behaviour. Mikians et al. [43] found evidence of price search discrimination using different virtual personas on Google. Hanak et al. [18] analyzed how search personalization on Google varied according to different demographics (e.g. age, gender), browsing history and geolocation, and found that only browsing history and geolocation significantly affected personalization of results. A follow-up study extended the work of Hannak et al. [18] to assess the impact of location, finding that personalization of results grows as physical distance increases [19]. Haim et al. [24] performed experiments to examine if suicide-related queries will lead to a "filter-bubble"; instead, they found that the decision to present Google's suicide-prevention result (SPR, with country-specific helpline information) was arbitrary (but persistent over time). In a follow up study, Scherr et al. [44] showed profound differences in the presence of the SPR between countries, languages and different search types (e.g. celebrity-suicide-related searches). Recently, virtual agents were used to measure the effects of randomization and differences on non-personalized results between search engines for the "coronavirus" query in different languages [25] and the 2020 U.S. Presidential Primary Elections [11].

News, images and video search results have also been subject to virtual agent-based auditing. Cozza et al. [13] found personalization effects for the recommendation section of Google News, but not for the general news section. In line with this, Haim et al. [14] found that only 2.5% of the overall sample of Google news results (N=1200) were exclusive to four constructed agents based on archetypical life standards and media usage. Image search results have also been audited for queries related to migrant groups [45], mass atrocities [31] and artificial intelligence [22]. A video search audit found that results are concentrated on YouTube for five different search engines [29], and the predominance of YouTube in the Google video carousel [46, 47]. Directly analyzing YouTube search results and Top 5 and Up-Next recommendations, Hussein et al. [30] showed personalization and "filter bubble" effects for misinformation topics after agents had developed a watch history on the platform.

Apart from search engine results, virtual agent auditing was used to study gender, race and browsing history biases in news and Google advertising [36, 48], price discrimination in e-commerce, hotel reservation and car rental websites [49, 50], music personalization in Spotify [51, 52], and news recommendations in the New York Times [53]. To our knowledge, four previous works have provided their programming code to facilitate data collection [18, 24, 35, 36]. Two of these programming solutions are built on top of the PhantomJS framework whose development has been suspended [18, 24]. Adfisher [36] specializes exclusively in Google Ads, and includes the automatic configuration of demographics information for the Google account, as well as statistical tests to find differences between groups. Haim [35] has provided a toolkit to set up a virtual agent architecture; the approach is generic and the bots can be programmed with a list of commands to create "recipes" that target specific websites or services. We contribute to this set of solutions by providing the source code of our browser extension [54] which simulates the search navigation on up to eight different search engines, including text, news, images, and video categories.

**Methodology**

The process of conducting algorithmic auditing, from our perspective, has two requirements: on one hand, the user information behaviour (e.g., browsing web search results) must be simulated appropriately; on the other hand, the data must be collected in a systematic way. Regarding the behaviour simulation, our methodology controls for factors that could affect the collection of web search results, so that they are comparable within and across search engines. We focus on the use case of a "default user" that attempts to browse anonymously, i.e. avoids personalization effects by removing historical data (e.g. cookies), but still behaves close to the way a human would do when using a browser (e.g. clicking and scrolling search pages). At the same time, we attempt to keep this behaviour consistent across several search engines, e.g. by synchronizing the requests. Effectively, the browsing behaviour is encapsulated in a browser extension, called WebBot [54].

For data collection, we have been using a distributed cloud-based research infrastructure. For each collection, we have configured a number of machines that vary depending on the experimental design. On each machine (2CPUs, 4GB RAM, 20GB HD), we installed CentOS and two browsers (Firefox and Chrome). In each browser, we installed two extensions: the WebBot that we briefly introduced above, and the WebTrack [55]. The tracker collects HTML from each page that is visited in the browser, and sends it to a server (16CPUs, 48GB RAM, 5TB HD), a different machine where all the content is stored, and where we can monitor the activity of the agents. In this section, we used the term virtual agent (or simply "agent") to refer to a browser that has the two extensions installed and that is configured for one of our collections.

A virtual agent in our methodology consists of an automated browser (through two extensions) that navigates through the search results of a list of query terms on a set of search engines, and that sends all the HTML of the visited pages to a server where the data is collected. The agent is initialized by assigning to it (1) a search engine and (2) the first item of the query terms list. Given that pair, the agent simulates the routine of a standard user performing a search on the following search categories of the engine: text, news, images and video. After that, it will simultaneously shift the search engine and the query term in each iteration to form the next pair and repeat the routine. This rest of this section describes the latest major version (version 3.x) of the browser extension that simulates the user behaviour, and later we will list differences in the older versions that have methodological implications.

The extension can be installed in Firefox and Chrome. Upon installation, the bot cleans the browser by removing all relevant historical data (e.g. cookies, local storage). For this, the extension requires the "browsingData" privilege. Table 1 presents the full lists of data types that are removed for Firefox and Chrome. After this, the bot downloads the lists of search engines and query terms that are previously defined as part of an experimental design (see Data Collections section).

| Browser | Data Types |
|---|---|
| Chrome | appcache, cache, cacheStorage, **cookies**, fileSystems, formData, history, indexedDB, **localStorage**, pluginData, passwords, serviceWorkers, webSQL |
| Firefox | cache, **cookies**, formData, history, indexedDB, **localStorage**, pluginData, passwords |

**Table 1. Data types that are cleaned during the installation and after each query.** The lists differ due to the differences between the browsers. A description of data is available for Chrome [56] and Firefox [57]. The bolded elements were included in version 3.0.

The navigation in the browser extension is triggered on the next exact minute (i.e., "minute o' clock") after a browser tab lands on the main page of any of the supported search engines: Google, Bing, DuckDuckGo, Yahoo, Yandex, Baidu, Sogou and So. Once triggered, the extension will use the first query term to navigate over the search result pages of the search engine categories (text results, news, images and videos). After each search routine, the browser is cleaned again according to Table 1. Table 2 briefly describes the general steps in the routines for each of the search engines.

|  | Text Results | News | Images | Videos |
|---|---|---|---|---|
| **Google** | Navigate 5 RPs. | Navigate 5 RPs. | Scroll and load 3 RPSs. | Navigate 5 RPs. |
| **Bing** | Navigate 7 or 8 RPs. | Scroll and load 10RPs. | Scroll and load 10 RPSs. | Scroll and load 14RPs. |
| **DDG** | Scroll and load 3 RPSs. | Scroll and load 3 RPSs. | Scroll and load 4 RPSs. | Scroll and load 3 RPSs. |
| **Yahoo** | Navigate 5 RPs. | Navigate 5 RPs | Scroll and load 5 RPSs. Click on load more images in each scroll. | Scroll and load 3 RPSs. Click on load more videos in each scroll. |
| **Yandex** | Navigate 1 RPs. | Navigate 1RPs. | Scroll and load 3 RPSs. | Scroll and load 3 RPSs. Click on load more videos in each scroll. |
| **Baidu** | Navigate 5 RPs. | Navigate 5 RPs. | Scroll and load 7 RPSs. | Scroll and load 7 RPSs. |
| **Sogou** | Navigate 5 RPs. | Not implemented | Scroll and load 6 RPSs. | Navigate 5 RPs. *URL bar is modified using push state.* |
| **So** | Navigate 5 RPs. *Wait after scrolling to the bottom, so ads and a navigation bar load.* | Not implemented | Scroll and load 10 RPSs. | Navigate 5 RPs. *(automatically redirects to video.360kan.com)* |

**Table 2. Summary of virtual agent routines for different search engines.** The table shows the search engine (first column), and the simulated actions for each search category, which are always visited in the order of column: text, news, images, videos. The bot always scrolls down to the bottom of each *Result Page* (RP), and each *Result Page Section* (RPS). RPSs refer to the content that is dynamically loaded on "continuous" search result pages (e.g. image search on Google). The bot uses Javascript to simulate the keyboard, click and scroll inputs, potentially triggering events that are part of the search engine internal code. Important particularities of search engines are highlighted in italics.

The search routines pursue two goals: first, to collect 50 results on each search category and, second, to keep the navigation consistent. For the most part, we succeed in reaching the first goal with the majority of search engines providing the required number of results. The only exception was Yandex, for which we decided to only collect the first page for text and news results because Yandex allows a very low number of requests per IP. After the limit is exceeded, Yandex detects the agent and blocks its activity by means of captchas. Our second goal was only fulfilled partially, because it is impossible to reach full consistency given multiple differences between search engines such as the number of results per page, speed of retrieval, the navigation mechanics (pagination, continuous scrolling, or scroll and click to load more), and other features highlighted in italics in Table 2.

To make the behaviour of agents more consistent, we tried to keep the search routines under 4 minutes and guaranteed that each search routine started at the same time by initializing a new routine every 7 minutes (with negligible differences due to internal machine clock differences). Additionally, the extension is tolerant to network failures (or slow server responses), because it refreshes a page that is taking too long to load (maximum of 5 attempts per result section). In the worst -case scenario, after 6.25 minutes an internal check is made to make sure that the bot is ready for the next iteration, i.e. the browser has been cleaned and landed in the corresponding search engine starting page, ready for the trigger of the next query term (that happens every 7 minutes).

To give a clearer idea of the agent functionality, Table 3 presents a detailed step by step description of the search routine implementation for an agent configured to start with the Google search engine (followed by Bing) in the Chrome navigator. The description assumes that the routine is automatically triggered by a terminal script, for example using Linux commands such as "crontab" or "at".

| # | Step description |
|---|---|
| 1 | A terminal script opens Chrome and waits 15 secs to execute Step 3. |
| 2 | The bot removes all historical data (see Table 1), downloads the query term and search engine lists from the server, and sets the first item in the query term list as the *current query term*. |
| 3 | The terminal script in Step 1 opens a tab in the browser with the https://google.com URL. |
| 4 | Upon landing on the search engine page, the bot resolves the consent agreement that pops up on the main page. |
| 5 | In the next exact "minute o'clock", the first search routine is triggered for the *current query term*. |
| 6 | The bot "types" the current query term in the input field of the main page of the search engine. Once the |
| | *current query term* is typed, the search button is clicked. |
| 7 | Once the text result page appears, we simulate the scroll down in the browser until the end of the page is reached. |
| 8 | If the bot has not reached the 5th result page, the bot clicks on the next page button and repeats step 7. Otherwise, it continues to step 9. |
| 9 | The bot clicks on the news search link, and repeats the behaviour used for text results (steps 7 and 8) |
| 10 | The bot clicks on the image search link and scrolls until the end of the page is reached. When the end of the page is reached, the bot waits for more images to load and then continues scrolling down. It repeats the process of scrolling and loading more images three times. |
| 11 | The bot clicks on the video search link, and repeats the behaviour used for text results (steps 7 and 8) |
| 12 | The bot navigates to a dummy page hosted at http://localhost:8000. Upon landing, the bot updates internal counters of the extension and removes historical data as shown in Table 1. |
| 13 | The bot navigates to the next search engine main page according to the list downloaded on Step 2 (e.g., https://bing.com), and sets the next element of the query term list (or the first element if the *current query term* is the last of the list) as the *current query term*. |
| 14 | Upon landing on the search engine page, the bot resolves the consent agreement that pops up on the main page. |
| 15 | After 7 minutes have passed since entering the previous query, the next search routine is triggered and continues from Step 6 (adjusting step 7 to 11 according to the next search engine in Table 2, for example https://bing.com. |

**Table 3. Detailed navigation process for an agent starting search routine on Google.** Each row corresponds to a step of the search routine for Google. The first column enumerates the step and the second gives its description. The process includes the steps that correspond to the agent setup before the actual routine starts (Steps 1 to 5), and steps that correspond to the routine of the next search engine (Steps 13 to 15).

The description in this section only involves the steps for one machine (and one browser), which simultaneously shifts to the new search engine and the new query term after the routine ends. Assuming a list of search engines, say <e1, e2>, and a list of query terms, say <q1,q2,q3,q4>, and an agent that is initialized with the engine e1 and query q1, i.e. <e1, q1>, then, the procedure will only consider the pairs <e1,q1>, <e2,q2>, <e1,q3>, <e2,q4>, and exclude the combinations <e2,q1>, <e1,q2>, <e2,q3>, <e1,q4>. To obtain results for all the combination of engines and queries, the researcher can (1) manipulate the list so that all pairs are included, e.g., one possible solution would be to repeat the query term twice in the query list, i.e. <q1,q1,...,q4,q4>, or (2) to use as many machines as search engines, e.g. one that is initialized to e1 and another to e2. The second alternative is preferred, because it keeps the search results synchronized assuming that all machines are started at the same.

On Table 4, we report relevant changes in the WebBot versions that have been used for the data collection rounds (See Data Collections section). We only include differences that have methodological implications because they either (1) have the potential to affect the results returned by the search engine (because they imply changes in the browser navigation), or (2) affect the way in which experiments are designed and set up. We do not include changes related to bug fixes, minor improvements or updates, adjustment of timers, or increase of robustness in terms of network failure tolerance and unexpected engine behaviours.

| Version | Features that affect the methodology or data collection |
|---|---|
| **1.0** 26.02.20 | a. Local storage and cookies were removed from the extension front end using the access privileges of the search engine javascript<br>b. Browser cleaning was performed in the last page (video search page)<br>c. It only accepted Yahoo consent agreement<br>d. Included the collection of at least 50 text, image and video search results<br>e. Each bot kept navigating the same engine<br>f. The list of queries terms was hard coded in the extension<br>g. Infinite number of reloading attempts upon network issues |
| **1.1** 26.10.20 | a. Acceptance of Google and Yandex consent agreements was added **[1.0.c]** |
| **2.0** 02.11.20 | a. News search category was integrated for all engines except So and Sogou **[1.0.d]** |
| **3.0** 22.02.21 | a. Local storage and cookies are removed from the extension backend using the "browsingData" privileges **[1.0.a]**<br>b. The browser clearing is performed in a (new) dummy page hosted in our server **[1.0.b]**<br>c. Acceptance of Bing consent agreement was added **[1.1.a]**<br>d. The number of collected pages for Yandex text and news search is limited to the first page, and the navigation shift to image and video search is implemented when a captcha banner pops up **[1.0.d, 2.0.a]**<br>e. The bot iterates through a list of search engines, shifting to a new engine after the end of each 7-minute routine **[1.0.e]**<br>f. The list of query terms and search engines is downloaded from server **[1.0.f]**<br>g. The number of reloads is limited to 5 times per search category **[1.0.g]** |
| **3.1** 05.03.21 | [n/a] |
| **3.2** 15.03.21 | a. The browser clearing is performed in a dummy page hosted in the same machine (http://localhost:8000) (corresponding to Step 12 in Table 3) **[3.0.b]** |

**Table 4. Relevant features of WebBot versions.** The first column indicates the version and the date when it was released. The second column enumerates features that could have an effect on the data collection of search results or on the experimental design. All the features correspond to changes with respect to previous versions, except for the first row (version 1.0). In that case, the included features are the ones that change in the following versions (and that differ from the navigation described in Table 3). The value inside the brackets, e.g. [3.0.c], at the end of each row in the second column refers to the previous feature that is affected.

We also enumerate the list of relevant changes. To reference these changes in the current document, we use the following format of abbreviations to refer to individual features: [version.feature], e.g. [1.0.a] refers to the first feature (a) listed in the table for version 1.0. Below we briefly discuss the implications of these changes grouped into three categories:

**Browser cleaning.** In version 3.0, we modified the way in which the browsing data of the browser is cleaned [3.0.a], so that local storage and cookies are removed (also) from the extension backend (as indicated in Table 1). Before that, the local storage and cookies were removed from the extension front end [1.0.a], so it could only remove cookies and storage that were allocated by the search engine (due to browsing security policies). For our first version, we were forced to do so due to a technical issue. Cleaning the local storage or cookies from the backend also removed those data from all installed extensions in the browser (not only the browsing data corresponding to the webpages), including the WebTrack [55]. This session data was generated when the virtual agent was set up by manually entering a token which is pre-registered in the server. A proper fix involved a change that automatically assigned a generic token to each machine and, due to time constraints and methodological implications, such a change would have potentially affected the data consistency of collections done with versions previous to 3.0. As part of the fix, we included a visit to a dummy page [3.0.b], where the cleaning is now performed instead of doing the cleaning on the last visited page (i.e. last video search page) of the navigation routine [1.0.b].

**Cookie consent.** Regulation such as the European Union's ePrivacy Directive (together with the General Data Protection Regulation, GDPR) and the California Consumer Privacy Act (CCPA) forced platforms to include cookie statements asking for the user consent to store and read cookies from the browser as well as to process personal information collected through them. Since we have focused on non-personalized search results, we decided to ignore these banners in the first extension version - except for Yahoo!, where the search engine window was blocked unless the cookies were accepted [1.0.c]. By version 1.1 release, Google also started forcing cookie consent, so we integrated it for Google and Yandex [1.1.a], and later on for Bing [3.0.c].

**Improvements.** The final category includes improvements that affect the way (1) in which the data is collected, such as the inclusion of news search results [2.0.a] or the limitations to the number of result pages collected for Yandex [3.0.d], or (2) in which the researcher designs experiments, such as the iteration over search engines [3.0.e] and the automatic downloading of search engines and query term lists [3.0.f]. It also includes the imposed limit (of 5) to the attempts of reloading a page [3.0.g] which improves the robustness at the cost of consistency because it introduces a condition under which the navigation path could be different, e.g. by completely skipping the image search results. The change [3.2.a] corresponds to a minor improvement that was only included for consistency with the description provided in Table 3.

**Experiments**

Starting February 2020, we have been using the WebBot extension to collect data to explore a multitude of research questions related to, for example, search engine differences, browser and geo-localization effects, visual portrayals in the image results, and source concentration. In total, we have performed 15 data collections with diverse experimental designs that are summarized in Table 5. For each of the collections, we rented and configured browsers using virtual machines provided by Amazon Web Services (AWS) via Amazon

Elastic Compute Cloud (Amazon EC2). The procedure to configure the infrastructure varied depending on the experimental design and the version of the bot extension. Prior to version 3.0, we manually set up each machine to register the tracker extension with a token. Starting from version 3.0, we only needed to configure the number of machines corresponding to the number of unique search engines we selected for the experiment. It became possible by creating images of the machines and then cloning them across the different geographical Amazon regions that were of interest for the research questions.

Some of the collections (C, G, I in Table 5) replicate the earlier rounds using the same selection of search queries (see column Replicates) to check the stability of results and to perform longitudinal analyses. The 4th and 5th collections have a low number of queries but the experimental set up was designed to monitor the changes in search results during a short time frame, so multiple iterations were performed for the same set of query terms.

| ID | Date | Version | Replicates | Agents | Regions | Engines | Browsers | Iterations | Query terms |
|---|---|---|---|---|---|---|---|---|---|
| A | 2020-02-27 | 1.0 | - | 200 | 1 | 6 | 2 | 1 | 120 |
| B | 2020-02-29 | 1.0 | - | 200 | 1 | 7 | 2 | 1 | 85 |
| C | 2020-10-30 | 1.1 | ID A | 200 | 1 | 6 | 2 | 1 | 169 |
| D | 2020-11-03 | 2.0 | - | 240 | 2 | 5 | 2 | 80 | 9 |
| E | 2020-11-04 | 2.0 | - | 20 | 2 | 5 | 2 | 330 | 3 |
| F | 2021-03-03 | 3.0 | - | 480 | 4 | 6 | 2 | 1 | 64 |
| G | 2021-03-10 | 3.1 | ID F | 480 | 4 | 6 | 2 | 1 | 63 |
| H | 2021-03-12 | 3.1 | - | 360 | 3 | 6 | 2 | 1 | 41 |
| I | 2021-03-17 | 3.2 | ID H | 360 | 3 | 6 | 2 | 1 | 41 |
| J | 2021-03-18 | 3.2 | - | 720 | 6 | 6 | 2 | 1 | 196 |
| K | 2021-03-19 | 3.2 | - | 384(*) | 6 | 6 | 2 | 2 | 89 |
| L | 2021-03-20 | 3.2 | - | 360 | 3 | 6 | 2 | 1 | 163 |
| M | 2021-05-07 | 3.2 | ID L | 360 | 3 | 6 | 2 | 1 | 163 |
| N | 2021-05-07 | 3.2 | ID J,K | 384(*) | 6 | 6 | 2 | 1 | 285 |
| O | 2021-05-14 | 3.2 | - | 360 | 6 | 6 | 2 | 1 | 8 |

**Table 5. Experimental design of the data collections.** From left to right, each column corresponds to: (1) an identifier (*ID*) of the collection used in the text of this paper, (2) the *date* of the collection, (3) the *version* of the bot utilized for the collection, (4) if the collection *replicates* queries and experimental set up of a previous collection, then such a collection is identified in this column, (5) the number of *agents* that were used for the collection, (6) the number of geographical *regions* in which agents where deployed, (7) the number of search *engines* and (8) the number of browsers that were configured for each collection, (9) the number of times (*iterations*) that each query was performed, and (10) the number of *query terms* included in the collection. (*) We assigned 24 extra machines to one of the regions (São Paulo) as this Amazon region seemed less reliable in a previous experiment.

All the collections included the same six search engines (Baidu, Bing, DuckDuckGo Google, Yahoo! and Yandex), except the collection B which also included two extra Chinese engines, So and Sogou, which were important given the nature of the collection and the research questions. The collection rounds E and F excluded Yandex, because the platform was detecting too many requests coming from our IPs (see Results section).

To understand the robustness of the method and scale of the collections, we present the results of the data collections in terms of coverage, size, and effective size (response variables). Coverage is the proportion of agents that collected data in each experimental condition according to the agents expected. We estimate this value by counting the agents that successfully collected at least one result page under each experimental condition and dividing it by the number of agents assigned to that condition. Size is the space that each collection occupies on the server. We estimate this number by adding up the kilobytes of each file that was collected for the collection. Effective size is the space of the collection excluding extra pages that are not relevant for the collection, e.g. home or cookie consent pages but also pages collected due to a delay between the end of the experiment and turning off the machines.

To help future researchers in estimating data collections, we provide average sizes per combination of search engine and results section. For this calculation, we include only those query terms for which we obtained the exact number of pages that we aimed for. To show that these values are robust, we compare two size estimates for each collection: in-sample estimates, calculated using only the averages corresponding to the query terms of that collection, and out-sample estimates, using the averages corresponding to the query terms that are not included in that collection.

**Results**

Figure 1 presents the coverage for 3 experimental conditions (browser, search category, and search engine) for our different collections. In multiple cases, we achieved near-perfect coverage and consistently collected above 80% of results. However, there are some clear gaps that we explain and discuss below.

| | | | Main | | | | | | News | | | | | | Images | | | | | | Videos | | | | | |
|---|---|---|---|---|---|---|---|---|---|---|---|---|---|---|---|---|---|---|---|---|---|---|---|---|---|---|
| | | | Baidu | Bing | DDG | Google | Yahoo! | Yandex | Baidu | Bing | DDG | Google | Yahoo! | Yandex | Baidu | Bing | DDG | Google | Yahoo! | Yandex | Baidu | Bing | DDG | Google | Yahoo! | Yandex |
| A (v1.0) | Chrome | 1.0 | .92 | .97 | .94 | .95 | .65 | | | | | | | 1.0 | .92 | .98 | .94 | .95 | .62 | 1.0 | .92 | 1.0 | .94 | .95 | .62 |
| 27.02.20 | Firefox | .93 | .98 | .98 | 1.0 | .88 | .49 | | | | | | | .93 | .98 | 1.0 | 1.0 | .88 | .42 | .93 | .94 | 1.0 | 1.0 | .88 | .41 |
| B (v1.0) | Chrome | .85 | .68 | 1.0 | .51 | | .69 | | | | | | | .74 | .67 | 1.0 | .49 | | .66 | .74 | .67 | 1.0 | .5 | | .66 |
| 29.02.20 | Firefox | .93 | .65 | .83 | .74 | | .4 | | | | | | | .81 | .64 | .83 | .72 | | .32 | .81 | .64 | .83 | .72 | | .32 |
| C (v1.1) | Chrome | 1.0 | .98 | .88 | .3 | 1.0 | .05 | | | | | | | 1.0 | .98 | .88 | .29 | 1.0 | .01 | .8 | .98 | .88 | .25 | .99 | .01 |
| 30.10.20 | Firefox | .94 | .97 | 1.0 | .99 | 1.0 | .01 | | | | | | | .94 | .96 | 1.0 | .97 | 1.0 | .0 | .94 | .94 | 1.0 | .95 | .93 | .0 |
| D (v2.0) | Chrome | .76 | .85 | .84 | .89 | .89 | | .75 | .84 | .84 | .81 | .89 | | .48 | .83 | .84 | .78 | .89 | | .38 | .81 | .84 | .75 | .86 | |
| 03.11.20 | Firefox | .84 | .9 | .94 | .99 | 1.0 | | .83 | .9 | .94 | .98 | 1.0 | | .51 | .9 | .94 | .98 | 1.0 | | .5 | .89 | .93 | .98 | 1.0 | |
| E (v2.0) | Chrome | .56 | .87 | .45 | .98 | 1.0 | | .53 | .87 | .44 | .61 | 1.0 | | .32 | .87 | .4 | .43 | 1.0 | | .24 | .87 | .43 | .23 | 1.0 | |
| 04.11.20 | Firefox | .03 | .87 | 1.0 | .84 | .54 | | .03 | .87 | 1.0 | .79 | .54 | | .03 | .87 | 1.0 | .78 | .54 | | .03 | .87 | .99 | .77 | .54 | |
| F (v3.0) | Chrome | .95 | .88 | .94 | .95 | .95 | .65 | .94 | .87 | .94 | .94 | .95 | .01 | .91 | .87 | .94 | .94 | .95 | .94 | .9 | .87 | .94 | .94 | .95 | .93 |
| 03.03.21 | Firefox | .87 | .8 | .86 | .85 | .85 | .33 | .85 | .78 | .86 | .84 | .85 | .01 | .83 | .79 | .86 | .84 | .85 | .87 | .82 | .79 | .86 | .84 | .84 | .85 |
| G (v3.1) | Chrome | .64 | .63 | .63 | .65 | .64 | .41 | .64 | .63 | .63 | .64 | .64 | .04 | .6 | .63 | .63 | .64 | .64 | .63 | .6 | .63 | .63 | .64 | .64 | .63 |
| 10.03.21 | Firefox | .84 | .86 | .85 | .88 | .87 | .54 | .83 | .84 | .85 | .86 | .87 | .03 | .81 | .84 | .85 | .86 | .87 | .86 | .81 | .84 | .85 | .86 | .87 | .84 |
| H (v3.1) | Chrome | .93 | .93 | .92 | .93 | .94 | .64 | .9 | .92 | .92 | .93 | .93 | .04 | .89 | .92 | .92 | .93 | .93 | .93 | .89 | .92 | .92 | .93 | .93 | .93 |
| 12.03.21 | Firefox | .95 | .94 | .93 | .94 | .95 | .39 | .91 | .93 | .93 | .94 | .95 | .02 | .9 | .93 | .93 | .94 | .95 | .94 | .9 | .93 | .93 | .94 | .95 | .94 |
| I (v3.2) | Chrome | .98 | .98 | .98 | .98 | .98 | .71 | .96 | .98 | .98 | .98 | .98 | .71 | .93 | .98 | .98 | .98 | .98 | .98 | .93 | .98 | .98 | .98 | .98 | .98 |
| 17.03.21 | Firefox | .99 | .99 | 1.0 | .99 | .99 | .44 | .97 | .98 | 1.0 | .99 | .99 | .44 | .97 | .98 | 1.0 | .99 | .99 | .99 | .97 | .98 | 1.0 | .99 | .99 | .99 |
| J (v3.2) | Chrome | .96 | .96 | .95 | .96 | .96 | .68 | .81 | .78 | .94 | .94 | .94 | .67 | .47 | .28 | .82 | .15 | .2 | .45 | .72 | .57 | .37 | .73 | .34 | .42 |
| 18.03.21 | Firefox | .95 | .95 | .94 | .94 | .95 | .37 | .72 | .74 | .94 | .91 | .93 | .37 | .55 | .29 | .41 | .34 | .36 | .44 | .77 | .58 | .18 | .77 | .39 | .77 |
| K (v3.2) | Chrome | .96 | .96 | .96 | .96 | .97 | .71 | .96 | .95 | .96 | .95 | .96 | .71 | .85 | .95 | .96 | .95 | .96 | .96 | .85 | .92 | .96 | .95 | .95 | .95 |
| 19.03.21 | Firefox | .98 | .98 | .98 | .97 | .98 | .36 | .98 | .91 | .97 | .96 | .95 | .36 | .83 | .95 | .97 | .93 | .95 | .95 | .88 | .91 | .96 | .89 | .91 | .92 |
| L (v3.2) | Chrome | .99 | 1.0 | .99 | 1.0 | 1.0 | .67 | .94 | .96 | .99 | .96 | .98 | .67 | .86 | .96 | .99 | .96 | .98 | .98 | .85 | .96 | .99 | .96 | .98 | .97 |
| 20.03.21 | Firefox | .98 | .98 | .98 | .98 | .98 | .41 | .93 | .93 | .98 | .95 | .96 | .41 | .87 | .94 | .98 | .94 | .96 | .97 | .86 | .92 | .97 | .94 | .96 | .95 |
| O (v3.2) | Chrome | .97 | .97 | .97 | .97 | .97 | .61 | .93 | .94 | .97 | .96 | .95 | .61 | .77 | .94 | .97 | .96 | .95 | .96 | .77 | .94 | .97 | .96 | .95 | .96 |
| 07.05.21 | Firefox | .93 | .94 | .93 | .94 | .94 | .44 | .89 | .86 | .91 | .92 | .92 | .44 | .75 | .86 | .88 | .86 | .92 | .93 | .74 | .83 | .87 | .86 | .87 | .86 |
| P (v3.2) | Chrome | .97 | .98 | .97 | .98 | .98 | .56 | .96 | .95 | .97 | .97 | .97 | .56 | .77 | .95 | .97 | .95 | .96 | .96 | .76 | .95 | .97 | .96 | .96 | .95 |
| 08.05.21 | Firefox | .96 | .95 | .95 | .93 | .94 | .45 | .91 | .82 | .9 | .92 | .89 | .45 | .76 | .88 | .89 | .86 | .92 | .94 | .75 | .84 | .87 | .81 | .85 | .86 |
| Q (v3.2) | Chrome | .95 | .98 | .95 | .97 | .97 | .89 | .95 | .98 | .95 | .84 | .96 | .89 | .76 | .98 | .95 | .84 | .96 | .96 | .76 | .98 | .95 | .84 | .96 | .95 |
| 14.05.21 | Firefox | .92 | .93 | .94 | .94 | .95 | .81 | .92 | .9 | .94 | .8 | .95 | .81 | .73 | .86 | .83 | .77 | .9 | .93 | .7 | .86 | .83 | .77 | .9 | .88 |

**Figure 1. Coverage per browser, result type, search engine and collection.** The heatmap displays the coverage (values from 0 to 1) obtained for each of the collections and three experimental conditions: browser, result type and search engine. In the first two rows, the collection (together with the version of the extension used and the date) and the browser are presented, and, in the columns, the result type (text results, news, images or videos) and the six engines most often included in our experiments. Coverage values which are closer to 0 is coloured with red tones, the ones closer to .5 with yellow tones and the ones closer to 1 with blue tones. The gray color is used for missing values, i.e., for conditions that were not included in the experimental design. The coverage for so.com and sogou.com (only used for collection 19.02.21) ranged between .48 and .68, except for sogou.com in Chrome in which was between .19 and .23.

**Poor coverage for Yandex.** Yandex restricts the number of search queries that come from the same IP and after the limit is reached it starts prompting captchas [58]. After several tests, we found that Yandex only blocks text and news search results, but not image and video ones. Therefore, we improved our extension by making it jump to image and video search when a captcha was detected. We can see that the coverage for images and news was fixed after collection F (version 3.0). However, coverage for news was still poor (see collections G,H,I). So, we decided to only collect the top 10 results for Yandex (i.e. the first page of search results) for text and news search, which allowed us to improve the consistency of coverage at the cost of volume. We did not experience these issues in the last collection (Q) because it only included 8 queries

**Coverage gaps before v3.0.** Most of these gaps were due to various small programming errors that triggered under special circumstances (e.g. lack of results for queries in certain languages) combined with the lack of recovery mechanisms in the extension. We also noticed that Google detected our extension more often for Chrome than for Firefox (see collections B and C), which caused low coverage.

**Differences between the browsers.** Apart from Chrome-based agents being more often detected as bots by Google, we noticed that Chrome performed poorly when it did not have visual focus from the graphical user interface, for which the operating system gives more priority. This problem was clearly observed in collection G, so all subsequent collections kept the visual focus on Chrome, which allowed us to address this limitation.

**Specific problems with particular collections.** Collection J included 720 agents and exceeded our infrastructure capabilities; the bandwidth of our server was not sufficient to attend the uploading requests in time. This explains the progressive degradation between the text and the video search results. Collection E was very distinct as (1) it had very few machines (only one per region and engine) and (2) it took over 4 days (see information about iterations in Table 5). Therefore, one single machine that failed (and did not recover) would heavily affect the coverage for the rest of the iterations in this case.

The first row of Table 6 presents the total size of each of the collections, followed by the effective size, i.e., the size of the files that correspond to the page that are targeted by the collection. Overall, the effective size is 95.46% of the total size (1.19 out of 1.25 Terabytes(TB)). The remaining 4.54% percent are composed of extra pages that do not contain search results, including search engine home, captcha, cookie and dummy (see v3.2b, Table 4) pages, but also search results pages that were collected after the end of the experiment due to a delay when stopping the machines and the iteration over the query list (Step 13 of Table 3) and from unintended queries (due to search engine automatic corrections and completions, or encoding problems, see Table 7).

|  | A (v1.0) 27.02.20 | B (v1.0) 29.02.20 | C (v1.1) 30.10.20 | D (v2.0) 03.11.20 | E (v2.0) 04.11.20 | F (v3.0) 03.03.21 | G (v3.1) 10.03.21 | H (v3.1) 12.03.21 | I (v3.2) 17.03.21 | J (v3.2) 18.03.21 | K (v3.2) 19.03.21 | L (v3.2) 20.03.21 | O (v3.2) 07.05.21 | P (v3.2) 08.05.21 | Q (v3.2) 14.05.21 |
|---|---|---|---|---|---|---|---|---|---|---|---|---|---|---|---|
| **Full** | 45.2 | 23.9 | 48.8 | 109.5 | 32.5 | 62.2 | 51.9 | 32.0 | 33.7 | 187.4 | 148.8 | 123.0 | 112.1 | 230.7 | 6.5 |
| **Effective** | 43.5 | 22.8 | 46.7 | 105.8 | 31.7 | 58.8 | 49.8 | 30.4 | 32.5 | 178.8 | 143.6 | 119.2 | 107.7 | 214.8 | 5.5 |
| **Exact cases** | 36.8 | 13.8 | 44.5 | 94.7 | 23.1 | 56.7 | 48.0 | 28.7 | 30.7 | 159.4 | 138.8 | 113.2 | 101.8 | 205.7 | 5.4 |
| **In-sample** | 46.1 | 21.9 | 50.4 | 116.3 | 39.7 | 66.1 | 66.2 | 32.1 | 32.3 | 307.1 | 152.1 | 118.9 | 113.4 | 232.7 | 6.0 |
| **Out-of-sample** | 47.5 | 27.1 | 66.4 | 114.4 | 39.5 | 68.2 | 67.1 | 32.8 | 32.8 | 310.6 | 151.1 | 132.0 | 132.0 | 237.0 | 6.4 |

**Table 6. Size estimates (in Gigabytes, GB) of the data collections.** Each column corresponds to the data collections. From top to bottom -Full: size of all the files of the collection; Effective: size of files corresponding to the search pages targeted by the experimental design; Exact cases: size of files corresponding to search sections that had the exact number of expected pages for that section (e.g. 5 pages for Google news); In-sample: size estimate based on the average size of the query terms used in each of the collections; Out-of-sample: estimate of the size based on the average sizes of the search sections excluding the query terms corresponding to each of the collections.

Figure 2 presents the average sizes of the search sections of each of the engines. To avoid distorting the averages due to incomplete result section or reloads due to network problems, we calculated them using only those searches that are complete (exact cases dataset), i.e., those in which we obtained the exact number of expected pages for each search section. The third row of table 6 displays the size of the dataset that correspond to the exact cases dataset; 92.44% of the effective dataset (1.1TB out of 1.19TB).

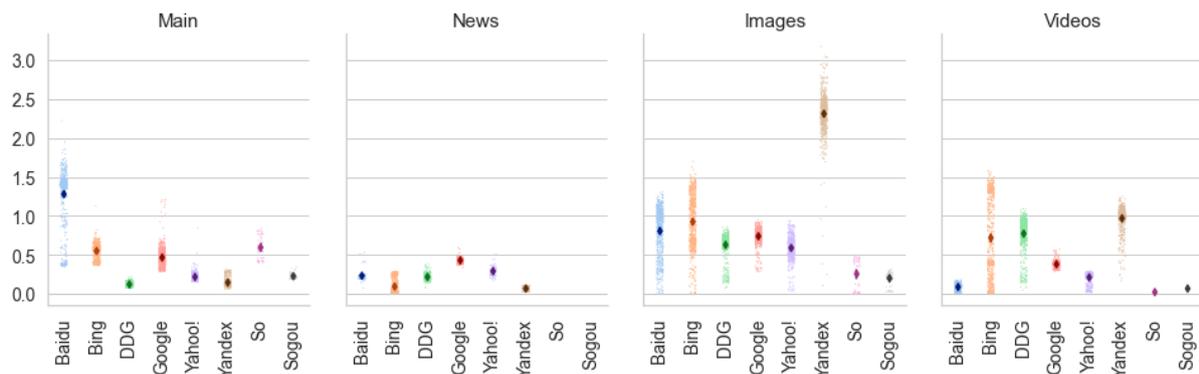

**Figure 2. Average size (in Megabytes) of the sections of each search engine.** From left to right, the plots present the average size in Megabytes (Y-axis) of the search sections (main results, news, images and videos). The X-axis shows the results per search engine. The dark dot indicates the average and includes 99% bootstrapped confidence intervals. All cases are plotted in lighter colour.

Although we only use exact cases for these calculations, there is an important variance in the size of each section (Figure 1), which stems from the query term and date of the collections. The variance is even bigger if all cases (not just the exact cases) are considered. To test if these averages would be useful to calculate future collections sizes, we estimate the size of the collection using in-sample and out-of-sample data (see Coverage and size in the Methodology). The estimates are displayed in the fourth and fifth rows of Table 6. In all cases (but one, collection I), the in-sample size estimate is higher than the effective size, and the out-of-sample estimates are close to the in-sample estimates. This indicates our averages are a good way to approximate the sizes of the collection.

**Discussion**

The coverage obtained with our method (Figure 1) demonstrates our incremental success in systematically collecting search engine results (RQ1a). In our most complete collection (I), our architecture simultaneously collected search outputs from 360 virtual agents totalling ~3172.92 pages every 7 minutes (811.52MB). This indicates that such architecture could be used for long-term search engine monitoring. An example of the use of such monitoring for the purpose of auditing is our collection round E, where we focused on three specific queries related to the US elections 2020 - Donald Trump, Joe Biden and US elections - for five days starting from the election day, in order to investigate potential bias in election outcome representation.

In terms of effective size (RQ1b), our method introduces very little noise to our collection, as most of the data (~95.45%) corresponds to results pages relevant for the queries of our experimental designs as opposed to extra pages not containing search results, e.g. cookie agreements, or unintending queries, e.g., due to engine automatic corrections or delays dismantling the infrastructure at the end of the experiment. Additionally, 92.44% of the effective size corresponds to complete queries, where the number of collected pages corresponds to the expected number according to the pagination of the search engine; thus, supporting our success in terms of coverage. We use this exact dataset to estimate sizes of search sections that are useful to calculate the size of future collections; using out of sample data, we provide evidence that our figures approximate the data collections sizes well.

Researchers should be aware of the complexities of collecting search engine results at a large scale with approaches like ours. This paper describes in detail all the steps we have taken to improve our methodology and Table 7 summarizes practical challenges that we had to address in this process (RQ2). We hope this will help researchers to succeed in their data collection endeavours.

| Category | Challenge |
|---|---|
| **Maintenance** | **Volatility of search engines layouts.** The HTML layout of search engines is in constant evolution making it practically impossible to develop an out-of-the-box solution, even if one limits the simulation of user behaviour to one platform. The changes are unannounced and unpredictable, so collections tools should be tested and adjusted before any new data collection. |
| | **Browser evolution.** Browsers change the way they organize and allow access to the different data types that they store, and it is necessary to keep the extension up to date. Browsers could offer more controls in relation to the host site in which the data (e.g. cookie) is added, and not only the third party that adds the cookie. |
| | **Cookie agreements.** A consequence of cleaning the browser data is that the behaviour must consider the acceptance of cookie statements of the different platforms each time a new search routine is started. Reginal differences are important as regulations differ. For example, the cookie statement no longer appeared for the machines with US-based IPs during our most recent data collection. |
| **Methodology** | **Disassociate IPs from search engines.** It is recommendable to let the agents iterate over the search engines, which brings three benefits: (1) avoid possible confounds between IPs and search results coming from the platforms, (2) decrease the number of search requests per IP to the same search engine which prevents the display of captcha pop-ups for most search engines, and (3) equally distribute the negative effects of the failure of one agent across all the search engines, so that the collection remains balanced. |
| **Failures** | **Network connectivity.** Although rare, network problems could cause major issues if not controlled appropriately. We included several contingencies to keep the machines synchronized, reduced data losses by resuming the procedure from predefined points (e.g., next query or next search section), and not saturating the server by allowing pauses between the different events. |
| | **Unexpected errors.** Multiple extrinsic factors can lead to browsers not starting properly or simply terminated. The underlying reasons for such failures are difficult to identify as all the machines are configured identically (clones), and we dismantle the architecture as soon as the collections are finished to save costs. |
| **Idiosyncrasies** | **Autocorrections and autocompletions**. Simulated browsing is not immune to being misled by corrections or completions that search engines offer, e.g. "protesta" (Spanish for protest) changed the query term to "protestant" due to the geolocation or "derechos LGBTQ" to "derechos LGBT" (which is problematic per se) |
| | **Character Encoding.** Certain search engines do not support characters of all languages, e.g. Baidu did not handle accents in Latin-based languages, e.g. the query "manifestação" was changed to "manifesta0400o". |

**Table 7. Practical challenges of search engine audits.** The first column organizes the practical challenges according to categories, and the second column explain the challenge in full, and includes examples and some recommendations.

Our methodology is the first to cover different search engines, and four different search categories. Although Google dominates web search market in the Western world, we also include other search engines, because some of them are slowly gaining market share (Bing, DuckDuckGo), whereas others keep historical relevance (Yahoo), or have considerable market shares in other countries (Yandex for Russia, and Baidu, Sogou and So for China). Ultimately, all search engine companies rank the same available information on the internet, but do it in a way that produces very different results [11, 25]. Given the difficulty of establishing baselines to properly evaluate whether some of these selections might be more or less skewed - or biased - towards certain interpretations of social reality [8], our approach addresses this limitation by offering the possibility of comparing results across different providers. With the combination of search engines, search categories, regions, and browsers, we can configure a wide range of conditions and adjust the experiments to target research questions related to the performance of these platforms in topics such as health, the elections, artificial intelligence and mass atrocity crimes.

We provide the code for the extension that simulates the user behaviour [54]. It is not as advanced as a recently released tool called ScrapeBot [35]; ScrapeBot is highly configurable, and offers an integrated solution for simulating user behaviours through "recipes" for collecting and extracting the data, as well as a web interface for configuring the experiments. Nonetheless, our approach holds some additional merit: first, by using the browser extensions API, we have full control of the HTML and the browser, which, for example, allows us to decide exactly when the browser data should be cleaned, and provides maximum flexibility in terms of interactions with the interface. Second, we collect all the HTML and not target specific parts of it to avoid potential errors when it comes to defining the specific selectors; once the HTML is collected a post-processing can be used to filter the desired parts. If one uses ScrapeBot, one is encouraged to target specific HTML sections, but it is also possible, and highly recommended based on our experience, to capture the full

HTML to avoid possible problems when the HTML of the services changes. Lastly, our approach clearly separates (1) the simulation of the browsing behaviour and (2) the collection of the HTML that is being navigated.

The latter allowed us to repurpose an existing tool which initially was aimed to be used for collection of human user data. Such an architecture enables more freedom in the use of each of the two components - namely, the WebBot and the WebTrack. Researchers could reuse a different bot, e.g. one that simulates browsing behaviour on a different platform, without worrying about changing the data collection architecture. Conversely, our WebBot could be used with a different web tracking solution to achieve similar results. A single caveat of the former scenario is our aggressive method to clean the browser history, which forced us to make modifications to the source code of the tracker that we used.

A limitation of virtual agent-based auditing approaches is that they depart from a simplified simulation of individual online behaviour. The user actions are simulated "robotically", i.e., the agent interacts with platforms in a scripted way; this is sufficient to collect the data, but not necessarily authentic. On one hand, it is possible that the way humans interact with pages (e.g., hovering the mouse for a prolonged time in a particular search result) have no effect on the search results, because these interactions are not considered by the platform algorithms. On the other hand, one cannot be certain until it is tested, given that the source code of the platforms is closed. Experiments that closely track user interactions with online platforms could help create more lifelike virtual agents. At the same time, it is important to revisit the differences in results obtained via the alternative ways of generating system outputs: simulating user behaviour via virtual agents, querying via platforms of APIs, and crowdsourcing data from real users.

Our browsing simulation approach is sufficient for experimental designs in which all machines follow a defined routine of searches, but, so far, the only possible variable that can be configured in each agent is the starting search engine, and even then, this is done manually in the start-up script (by preparing the number of machines corresponding to the number of unique search engines that are going to be included). A more sophisticated approach can allow more flexibility in configuring each virtual agent.

**Conclusion**

In this paper, we offer an overview over the process of setting up an infrastructure to systematically collect data from search engines. We document the challenges involved and improvements undertaken, so that future researchers can learn from our experiences. Despite challenges, we demonstrate the successful performance of our infrastructure and present evidence that algorithm audits are scalable. We conclude that virtual agents can be used for long-term algorithm auditing, for example to monitor long-lasting events, such as the current COVID-19 pandemic, or century affairs, such as climate change and human rights.

**Declaration of interest statement**

We have no conflicts of interest to disclose.

WORKING PAPER: SCALING UP SEARCH ENGINE AUDITS                    28## References

[1]  Gillespie T. The relevance of algorithms. *Media Technol Essays Commun Mater Soc* 2014; 167: 167.

[2]  Noble SU. *Algorithms of oppression: how search engines reinforce racism*. New York: New York University Press, 2018.

[3]  O'neil C. *Weapons of math destruction: How big data increases inequality and threatens democracy*. Crown, 2016.

[4]  Mittelstadt B. Automation, Algorithms, and Politics| Auditing for Transparency in Content Personalization Systems. *Int J Commun* 2016; 10: 12.

[5]  Bandy J. Problematic Machine Behavior: A Systematic Literature Review of Algorithm Audits. *ArXiv210204256 Cs*, http://arxiv.org/abs/2102.04256 (2021, accessed 23 April 2021).

[6]  Diakopoulos N, Trielli D, Stark J, et al. I Vote For—How Search Informs Our Choice of Candidate. In: *Digital Dominance: The Power of Google, Amazon, Facebook, and Apple*. Oxford University Press, 2018, p. 22.

[7]  Hu D, Jiang S, E. Robertson R, et al. Auditing the Partisanship of Google Search Snippets. In: *The World Wide Web Conference*. New York, NY, USA: Association for Computing Machinery, pp. 693–704.

[8]  Kulshrestha J, Eslami M, Messias J, et al. Quantifying Search Bias: Investigating Sources of Bias for Political Searches in Social Media. In: *Proceedings of the 2017 ACM Conference on Computer Supported Cooperative Work and Social Computing*. New York, NY, USA: Association for Computing Machinery, pp. 417–432.

[9]  Metaxa D, Park JS, Landay JA, et al. Search Media and Elections: A Longitudinal Investigation of Political Search Results. *Proc ACM Hum-Comput Interact* 2019; 3: 129:1-129:17.

[10] Trielli D, Diakopoulos N. Search as news curator: The role of Google in shaping attention to news information. In: *Proceedings of the 2019 CHI Conference on Human Factors in Computing Systems*. 2019, pp. 1–15.

[11] Urman A, Makhortykh M, Ulloa R. The Matter of Chance: Auditing Web Search Results Related to the 2020 U.S. Presidential Primary Elections Across Six Search Engines. *Soc Sci Comput Rev* 2021; 08944393211006863.

[12] Courtois C, Slechten L, Coenen L. Challenging Google Search filter bubbles in social and political information: Disconforming evidence from a digital methods case study. *Telemat Inform* 2018; 35: 2006–2015.

[13] Cozza V, Hoang VT, Petrocchi M, et al. Experimental Measures of News Personalization in Google News. In: Casteleyn S, Dolog P, Pautasso C (eds) *Current